\documentclass[aps,amsmath,amssymb,reprint,groupedaddress,floatfix]{revtex4-1}

\usepackage{graphicx}
\usepackage{floatrow}
\usepackage{bm}
\usepackage{amsmath}

\usepackage{amssymb}
\usepackage{color}
\usepackage{times}
\usepackage{float}
\usepackage[sort&compress]{natbib}
\usepackage[normalem]{ulem}

\usepackage{hyperref}

\hypersetup{
 colorlinks=true,
citecolor=black,
linkcolor=black,
urlcolor=black
 }

\usepackage[colorinlistoftodos,textsize=small]{todonotes}

\newcommand{\xtr}{x_{\mathrm{tr}}}
\newcommand{\T}{\mathcal{T}}

\begin{document}

\title{L\'evy flights and L\'evy walks under stochastic resetting}

\author{Bartosz \.Zbik}
\email{bartosz.zbik@student.uj.edu.pl}
\affiliation{Faculty of Physics, Astronomy and Applied Computer Science,
Jagiellonian University, \L{}ojasiewicza 11, 30-348 Krak\'ow, Poland}

\author{Bart{\l}omiej Dybiec}
\email{bartlomiej.dybiec@uj.edu.pl}
\affiliation{Institute of Theoretical Physics
and Mark Kac Center for Complex Systems Research,
Faculty of Physics, Astronomy and Applied Computer Science,
Jagiellonian University, \L{}ojasiewicza 11, 30-348 Krak\'ow, Poland}


\date{\today}

\begin{abstract}
Stochastic resetting is a protocol of starting anew, which can be used to facilitate the escape kinetics.
We demonstrate that restarting can accelerate the escape kinetics from a finite interval restricted by two absorbing boundaries also in the presence of heavy-tailed, L\'evy type, $\alpha$-stable noise.
However, the width of the domain where resetting is beneficial depends on the value of the stability index $\alpha$ determining power-law decay of jump length distribution.
For heavier  (smaller $\alpha$) distributions the domain becomes narrower in comparison to lighter tails.
Additionally, we explore connections between L\'evy flights and L\'evy walks in presence of stochastic resetting.
First of all, we show that for L\'evy walks, the stochastic resetting can be beneficial also in the domain where coefficient of variation is smaller than 1.
Moreover, we demonstrate that in the domain where LW are characterized by a finite mean jump duration/length, with the increasing width of the interval LW start to share similarities with LF under stochastic resetting.
\end{abstract}

\pacs{02.70.Tt,
 05.10.Ln, 
 05.40.Fb, 
 05.10.Gg, 
 02.50.-r, 
 }

%
\maketitle
%


%
%
\section{Introduction \label{sec:introduction}}
Since pioneering works of Smoluchowski \cite{smoluchowski1906b2}, Einstein \cite{einstein1905}, Langevin \cite{langevin1908}, Perrin \cite{perrin1909mouvement} and Kramers \cite{kramers1940} studies of Brownian motion and random phenomena attracts steadily growing interest.
Probabilistic explanation of properties of Brownian motion boosted development of the theory of stochastic processes \cite{vankampen1981,gardiner2009}, increased our understanding of random phenomena \cite{Zwanzig2001} and opened studies on noise driven systems \cite{horsthemke1984} and random walks \cite{montroll1984,metzler2000,metzler2004}.

The Wiener process (Brownian motion --- BM) is one of the simplest examples of continuous (time and space) random processes.
Its mathematical properties nicely explains observed properties of Brownian motion \cite{brown1828}, e.g., the linear scaling of the mean square displacement \cite{perrin1909mouvement,nordlund1914new}.
It can be extended in multiple ways, e.g., by assuming more general jump length distribution, introducing memory or assuming finite propagation velocity.
In that context, L\'evy flights (LF) \cite{dubkov2008,chechkin2008introduction} and L\'evy walks (LW) \cite{shlesinger1986,zaburdaev2015levy,denisov2004} are two archetypal types of random walks \cite{montroll1984}.
In the LF it is assumed that displacements are immediate and generated from a heavy-tailed, power-law distribution.
At the same time in the LW, a random walker travels with a finite velocity $v$ for random times distributed according to a power-law density.

The assumption that individual jump lengths follow a general $\alpha$-stable density is supported by multiple experimental observations demonstrating existence of more general than Gaussian fluctuations.
Heavy-tailed, power-law fluctuations have been observed in plenitude of experimental setups including, but not limited to,
biological systems \cite{bouchaud1991}, dispersal patterns of humans and animals \cite{brockmann2006,sims2008}, search strategies \cite{shlesinger1986,reynolds2009}, gaze dynamics \cite{amor2016}, balance control \cite{cabrera2004, collins1994random},
rotating flows \cite{solomon1993}, optical systems and materials \cite{barthelemy2008,mercadier2009levyflights}, laser cooling \cite{barkai2014}, disordered media \cite{bouchaud1990}, financial time series \cite{laherrere1998,mantegna2000,lera2018gross}.
Properties of systems displaying heavy-tailed, non-Gaussian fluctuations are studied both experimentally \cite{solomon1993,solomon1994,amor2016} and theoretically \cite{metzler2000,barkai2001,chechkin2006,jespersen1999,klages2008,dubkov2008,touchette2007,touchette2009,chechkin2009,dybiec2012,kusmierz2014}.
L\'evy flights attracted considerable attention due to their well-known mathematical properties, e.g., self similarity, infinite divisibility and generalized central limit theorem.
Therefore, the $\alpha$-stable noises are broadly applied in diverse models displaying anomalous fluctuations or describing anomalous diffusion.

Stochastic resetting \cite{evans2011diffusion,evans2020stochastic,gupta2022stochastic} is a protocol of starting anew, which can be applied (among others) to increase efficiency of search strategies.
In the simplest version, it assumes that the motion is started anew at random times, i.e., restarts are triggered temporally making times of starting over independent of state of the system, e.g., position.
Among multiple options, resets can be performed periodically (sharp resetting) \cite{pal2017first}, or at random time intervals following exponentiall (Poissonian resetting) \cite{evans2011diffusion} or a power-law \cite{nagar2016diffusion} density.
Starting anew can be also spatially induced \cite{dahlenburg2021stochastic}.
Escape kinetics under stochastic resetting display universal properties \cite{reuveni2016optimal,pal2017first} 
regarding relative fluctuations of first passage times as measured by the coefficient of variation ($CV$) {and as such can be also treated in the unified approach \cite{chechkin2018random}}.
{Typically it is assumed that the restarting is immediate and does not generate additional costs, however options with finite return velocity \cite{pal2019time,zhou2021gaussian}, overheads \cite{pal2020search,bodrova2020resetting,sunil2023cost} or soft (due to attracting force) resetting \cite{xu2022stochastic} are also explored.}

Stochastic resetting attracted considerable attention due to its strong connection with search strategies \cite{reynolds2009,viswanathan2011physics,palyulin2014levy} {and its ability of reducing the spread of particles \cite{pal2015diffusion,zhou2020continuous,sandev2022heterogeneous} even in the case of partial resetting \cite{di2023time}}.
During the search an individual/animal is interested in minimization of the time needed to find a target, which in turn is related to the first passage problem \cite{redner2001}.
In setups where due to long excursions in the wrong direction, i.e., to points distant from the target \cite{kusmierz2014firstorder,mendez2021ctrw,kusmierz2015optimal,stanislavsky2021optimal},
the mean first passage time (MFPT) can diverge.
In such cases, stochastic resetting is capable of turning the MFPT finite.
Furthermore, it can optimize already finite MFPT \cite{reuveni2016optimal,pal2017first}.
Stochastic resetting is capable of minimization of the time to find a target when the coefficient of variation
(the ratio between the standard deviation of the first passage times and the MFPT in the absence of
stochastic resetting)
is greater than unity \cite{reuveni2016optimal,pal2017first}.

Not surprisingly, the stochastic resetting is capable of minimizing the MFPT from a finite interval restricted by two absorbing boundaries \cite{pal2019first}.
As demonstrated in \cite{dybiec2017levy}, in the case of escape from finite intervals restricted by two absorbing boundaries mean first passage time for L\'evy flights and L\'evy walks display similar scaling \cite{dybiec2017levy} as a function of the interval width.
Therefore, one can study properties of escape from finite intervals under combined action of L\'evy noise and stochastic resetting with special attention to verification if properties of escape kinetics still bears some similarities with LWs under restarts.

The model under study is described in the next section (Sec.~\ref{sec:model} Model and Results).
Sec.~\ref{sec:levywalks} (L\'evy walks under stochastic resetting) analyzes properties of LW on finite intervals under stochastic resetting and compares them with properties of corresponding LF.
The manuscript is closed with Summary and Conclusions (Sec.~\ref{sec:summary}).

%
%
\section{Model and results \label{sec:model}}

The noise driven escape (from any domain of motion $\Omega$) is a stochastic process, therefore individual first passage times are not fixed but random.
For first passage times it is possible to calculate --- the relative standard deviation --- the coefficient of variation (CV) \cite{pal2017first}
\begin{equation}
 CV= \frac{\sigma( t_{\mathrm{fp}}) }{ \langle t_{\mathrm{fp}} \rangle} = \frac{\sigma( t_{\mathrm{fp}})}{\mathcal{T}}=\frac{\sqrt{\mathcal{T}_2-\mathcal{T}^2}}{\mathcal{T}} = \sqrt{ \frac{\mathcal{T}_2}{\mathcal{T}^2} -1},
 \label{eq:cv}
\end{equation}
which is the ratio between the standard deviation $\sigma(t_{\mathrm{fp}})$ of the first passage times (FPT) $t_{\mathrm{fp}}$ and the mean first passage time $\mathcal{T}=\langle t_{\mathrm{fp}} \rangle$.
In addition to statistical applications, the coefficient of variation plays a special role in the theory of stochastic resetting \cite{reuveni2016optimal,pal2017first,pal2019first}.
It provides a useful universal tool for assessing potential effectiveness of stochastic restarting which can be used to explore various types of setups under very general conditions.
Typically stochastic resetting can facilitate the escape kinetics in the domain where $CV>1$.
Therefore, examination of CV given by Eq.~(\ref{eq:cv}) (constrained by the fact that resets are performed to the same point from which the motion was started) can be a starting point for exploration of effectiveness of stochastic resetting.

The escape of a free particle from a finite interval $(-L,L)$ under action of L\'evy noise, i.e., escape of $\alpha$-stable, L\'evy type process, can be characterized by the mean first passage time (MFPT) $\mathcal{T}$ which reads \cite{getoor1961}
\begin{equation}
\mathcal{T}(x_0)= \frac{(L^2-|x_0|^2)^{\alpha/2}}{\Gamma(1+\alpha) \sigma^\alpha} 
= \frac{(1-|x_0/L|^2)^{\alpha/2}}{\Gamma(1+\alpha) } \times  \frac{L^\alpha}{\sigma^\alpha} 
\label{eq:mfpt}
\end{equation}
and the second moment $\mathcal{T}_2=\langle t_{\mathrm{fp}}^2 \rangle$ given by \cite{getoor1961}
\begin{eqnarray}
\label{eq:t2}
\mathcal{T}_2(x_0) & = & \frac{\alpha L^\alpha}{\left[ \Gamma(1+\alpha)\sigma^\alpha \right]^2} \\ \nonumber
& \times & \int_{|x_0|^2}^{L^2} \left[ t-|x_0|^2 \right]^{\frac{\alpha}{2}-1} {}_2F_1\left[ -\frac{\alpha}{2},\frac{1}{2},\frac{1+\alpha}{2}; \frac{t}{L^2} \right]dt,
\end{eqnarray}
where $x_0$ is the initial condition, ${}_2F_1(a,b;c;z)$ stands for the hypergeometric function, while $\Gamma(\dots)$ is the Euler gamma function.
From Eqs.~(\ref{eq:mfpt}) and (\ref{eq:t2}) with $\alpha=2$ one gets
\begin{equation}
    CV = \sqrt{\frac{2}{3}} \sqrt{\frac{L^2+x_0^2}{L^2-x_0^2}}
    = \sqrt{\frac{2}{3}} \sqrt{\frac{1+(x_0/L)^2}{1-(x_0/L)^2}}.
    \label{eq:cvalpha2}
\end{equation}
As it implies from Eqs.~(\ref{eq:mfpt}) and~(\ref{eq:t2}), the coefficient of variation does not depend on the scale parameter $\sigma$.
The independence of the CV on the scale parameter $\sigma$ can be intuitively explained by the fact that $\sigma$ can be canceled by time rescaling.
Such a transformation (linearly) rescales individual FPTs and consequently in exactly the same way the MFPT and  the standard deviation making their ratio $\sigma$ independent.
From Eq.~(\ref{eq:cvalpha2}) it implies that $CV>1$ for
\begin{equation}
    x_0 \in \left(-L,-\frac{L}{\sqrt{5}} \right) \cup \left(\frac{L}{\sqrt{5}},L \right)
    \label{eq:res-cond}
\end{equation}
what is in accordance with earlier findings \cite{pal2019first}.

Equivalently, the setup corresponding to Eqs.~(\ref{eq:mfpt}) and~(\ref{eq:t2}) can be described by the Langevin equation
\begin{equation}
    \frac{dx}{dt} = \xi_\alpha(t)
    \label{eq:langevin}
\end{equation}
and studied by methods of stochastic dynamics.
In Eq.~(\ref{eq:langevin}), the $\xi_\alpha$ is the symmetric $\alpha$-stable L\'evy type noise and $x(t)$ represents the particle position (with the initial condition $x(0)=x_0$).
The $\alpha$-stable noise is a generalization of the Gaussian white noise to the nonequilibrium realms \cite{janicki1994}, which for $\alpha=2$ reduces to the standard Gaussian white noise.
The symmetric $\alpha$-stable noise is related to the symmetric $\alpha$-stable process $L(t)$, see Refs.~ \cite{janicki1994,dubkov2008}.
Increments $\Delta L=L(t+\Delta t)-L(t)$ of the $\alpha$-stable process are independent and identically distributed random variables following an $\alpha$-stable density with the characteristic function \cite{samorodnitsky1994,janicki1994}
\begin{equation}
\varphi(k) = \langle \exp(i k \Delta L ) \rangle = \exp\left[ - \Delta t \sigma^{\alpha}\vert k\vert^{\alpha} \right].
    \label{eq:levycf}
\end{equation}
Symmetric $\alpha$-stable densities are unimodal probability densities defined by the characteristic function with probability densities given by elementary functions only in a limited number of cases ($\alpha=1$ Cauchy density, $\alpha=2$ Gauss distribution), however in more general cases can be expressed using special functions \cite{gorska2011}.
The stability index $\alpha$ ($0<\alpha \leqslant 2$) determines the tail of the distribution, which for $\alpha<2$ is of power-law type $p(x) \propto |x|^{-(\alpha+1)}$.
The scale parameter $\sigma$ ($\sigma>0$) controls the width of the distribution, which can be characterized by an interquantile width or by fractional moments $\langle |x|^\kappa \rangle$ of order $\kappa$ ($0<\kappa<\alpha$), because the $\alpha$-stable variables with $\alpha<2$ cannot be quantified by the variance which diverges.
Within studies we set the scale parameter to unity, i.e., $\sigma=1$.

The MFPT can be calculated from multiple trajectories generated according to Eq.~(\ref{eq:langevin}) as the average of the first passage times
\begin{equation}
 \mathcal{T}(x_0) = \langle t_{\mathrm{fp}} \rangle =
	 \langle \inf\{t : x(0)=x_0 \;\land\; |x(t)| \geqslant L \} \rangle,
	 \label{eq:mfptdef}
\end{equation}
while $\mathcal{T}_2(x_0)$ is the second moment of the first passage time.
The Langevin equation~(\ref{eq:langevin}) can be approximated with the (stochastic) Euler--Maruyama method \cite{higham2001algorithmic,mannella2002}
\begin{equation}
x(t+\Delta t) = x(t) + \xi_\alpha^t \Delta t^{1/\alpha},
    \label{eq:integration}
\end{equation}
where $\xi_\alpha^t$ represents a sequence of independent identically distributed $\alpha$-stable random variables \cite{chambers1976,weron1995,weron1996}, see Eq.~(\ref{eq:levycf}).

\begin{figure}[!h]
    \centering
    \includegraphics[width=0.8\columnwidth]{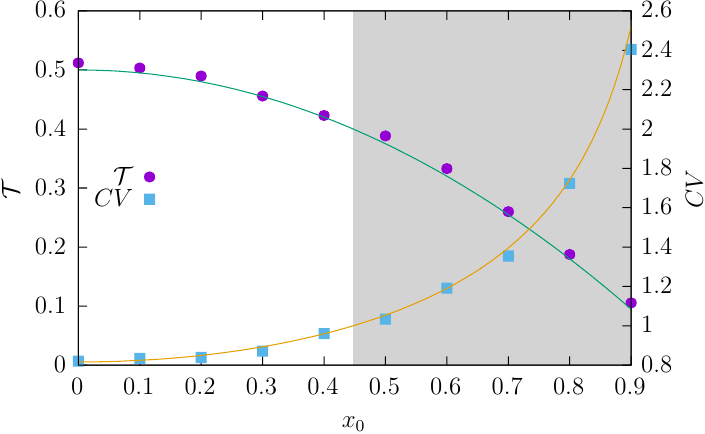}
    \caption{Numerically estimated MFPT (dots) and $CV$ (squares) as a function of the initial condition $x_0$ for $\alpha=2$ along with theoretical values (solid lines).
    The grayed region shows the domain where $CV(x_0)>1$.}
    \label{fig:cvalpha2}
\end{figure}

For the model under study, the coefficient of variation $CV(x_0)$ is a symmetric function of the initial condition $x_0$, i.e., $CV(x_0)=CV(-x_0)$ as the escape problem (due to noise symmetry) is symmetric with respect of the sign change in the initial condition $x_0$.
The symmetry implies from the system symmetry (symmetric boundaries and symmetric noise) and consequently is visible not only in Eq.~(\ref{eq:cvalpha2}) but also in Eqs.~(\ref{eq:mfpt}) -- (\ref{eq:t2}).
The condition $CV>1$ is a sufficient, but not necessary condition, when stochastic resetting can facilitate the escape kinetics \cite{rotbart2015michaelis,pal2019landau,pal2022inspection}.
As it implies from Eq.~(\ref{eq:res-cond}), stochastic resetting can accelerate the escape kinetics if initial condition, which is equivalent to the point from which the motion is restarted, sufficiently breaks the system symmetry, i.e., if restarting the motion anew is more efficient in bringing a particle towards the target (edges of the interval)  than waiting for a particle to approach the target (borders).
{In other words, for the escape from a finite interval, the point from which the motion is restarted is sufficiently far from the center of the interval ($x=0$).}
Fig.~\ref{fig:cvalpha2} compares results of numerical simulations (points) with theoretical predictions (solid lines) for $\alpha=2$ (Gaussian white noise driving) demonstrating perfect agreement.
Due to system symmetry, in Fig.~\ref{fig:cvalpha2}, we show results for $x_0>0$ only.
{Moreover, we set the interval half-width $L$ to $L=1$ and the scale parameter $\sigma$ to $\sigma=1$, what is equivalent to the tranformation of the escape problem to the dimensionless variables, see Appendix~\ref{sec:units}.}

\begin{figure}[!h]
    \centering
    \includegraphics[width=0.8\columnwidth]{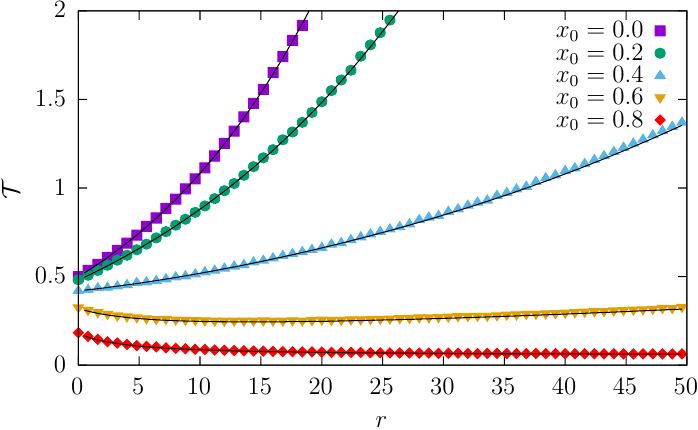}
    \caption{MFPT under stochastic (Poissonian) restarting for the GWN ($\alpha=2$) driving as a function of the resetting rate $r$.
 Various lines correspond to different initial conditions $x_0$.
 	Points represent results of numerical simulations while solid lines theoretical predictions given by Eq.~(\ref{eq:mfptreset}).}
    \label{fig:mfpta2}
\end{figure}

Stochastic resetting, i.e., starting anew from the initial conditions $x_0$ can be used to facilitate the escape kinetics.
One of common restarting schemes, is the so-called fixed rate (Poissonian) resetting, for which the distribution of time intervals between two consecutive resets follow the exponential density $\phi(t) = r \exp(-rt)$, where $r$ is the (fixed) reset rate.
Thus, the mean time between two consecutive restarts reads $\langle t \rangle = 1 / r$.
The MFPT under Poissonian resetting for a process driven by GWN $(\alpha = 2)$ \cite{pal2019first} from the $(-L,L)$ interval restricted by two absorbing boundaries reads
\begin{equation}
\mathcal{T}(x_0) = \frac{1}{r} \left[ \frac{  \sinh  \frac{2L}{\sqrt{  \sigma^2/r}}   }{    \sinh \frac{L-x_0}{\sqrt{ \sigma^2/r}}   +\sinh   \frac{x_0+L}{\sqrt{  \sigma^2/r}}  } -1 \right].
    \label{eq:mfptreset}
\end{equation}
Fig.~\ref{fig:mfpta2} presents MFPT as a function of resetting rate $r$ for various initial positions $x_0$.
MFPTs have been estimated using the so-called direct approach \cite{pal2017first}.
Within such a scheme from the simulation of the system without resets the (unknown for $\alpha<2$) first passage time density is estimated.
In the next step, instead of simulating the Langevin dynamics under stochastic resetting, pairs of first passages times (in the absence of resetting) and resetting times are generated, until the first passage time is smaller than the resetting time.
The first passage time under restart is equal to the sum of (all) generated time intervals between resets increased by the last first passage time, see \cite[Fig.~4($a$)]{pal2017first}.
In Fig.~\ref{fig:mfpta2} points representing results of computer simulations with $\alpha=2$ nicely follow solid lines demonstrating theoretical predictions given by Eq.~(\ref{eq:mfptreset}).

\begin{figure}[!h]
    \centering
    \includegraphics[width=0.8\columnwidth]{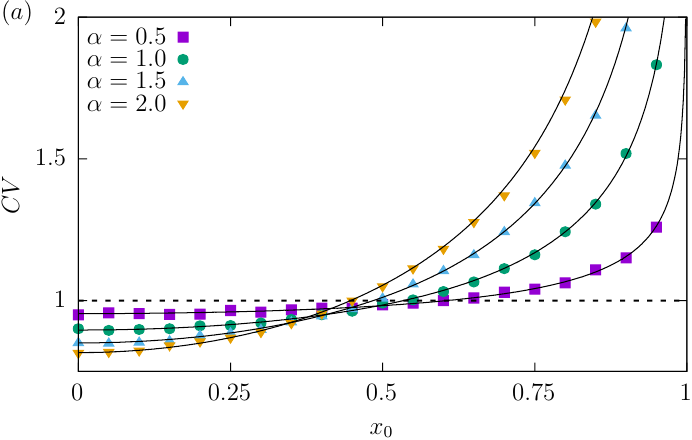}
    \includegraphics[width=0.8\columnwidth]{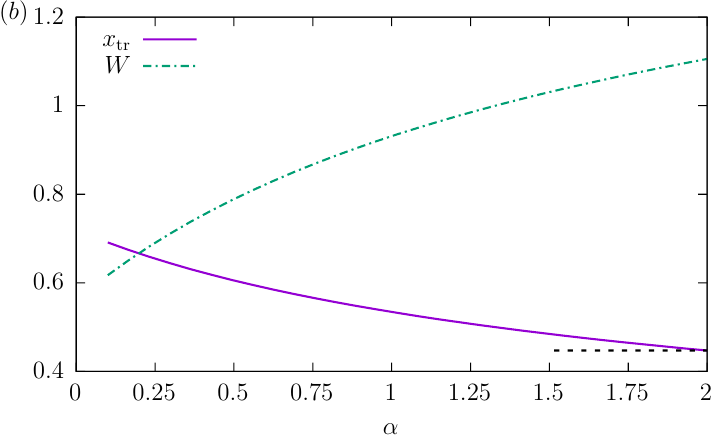}
    \caption{Top panel ($a$): numerically estimated $CV(x_0)$ for $\alpha\in\{0.5,1.0,1.5,2.0\}$ with $L=1$ (points) complemented by theoretical results (solid lines), see Eqs.~(\ref{eq:cv}) -- (\ref{eq:t2}).
    The dashed line in the top panel shows $CV=1$ level.
    Bottom panel ($b$): numerically calculated $\xtr(\alpha)$ such that $CV(\xtr)=1$ (solid line) and the width $W(x_0)$ of the domain where resetting is beneficial (dot-dashed line), see Eq.~(\ref{eq:width}). For $x_0>\xtr$ the coefficient of variation is greater than 1. The dashed line indicates the Gaussian white noise limit, i.e., $\xtr=1/\sqrt{5}$.}
    \label{fig:cvalpha}
\end{figure}

\begin{figure}[!h]
    \centering
    \includegraphics[width=0.8\columnwidth]{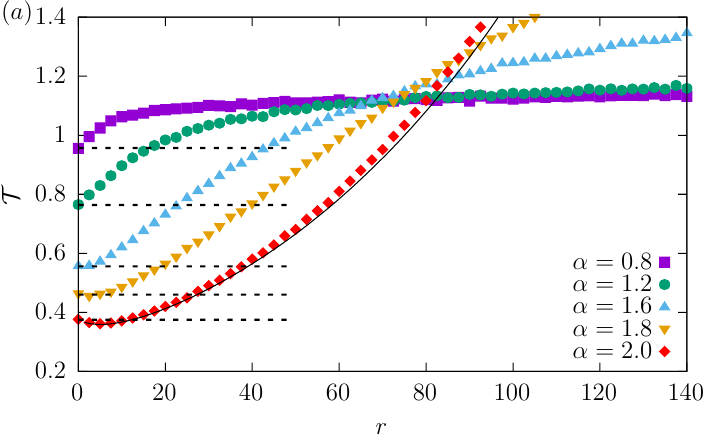}
	\includegraphics[width=0.8\columnwidth]{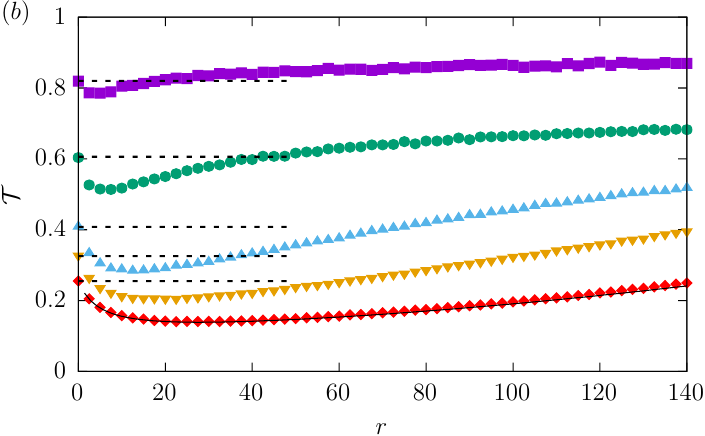}
    \caption{MFPT under Poissonian resetting. Various points represent different values of the stability index $\alpha$.
	The solid line depicts theoretical dependence for $\alpha=2$, see Eq.~(\ref{eq:mfptreset}), while dashed lines $\mathcal{T}(x_0,r=0)$, see Eq.~(\ref{eq:mfpt}).
 In the top panel ($a$) the initial condition is set to $x_0=0.5$ while in the bottom panel ($b$) to $x_0=0.7$.
 }
    \label{fig:mfptreset-lf}
\end{figure}

With help of Eqs.~(\ref{eq:mfpt}) -- (\ref{eq:t2}) more general driving than the Gaussian white noise can be studied.
As it is already visible from Fig.~\ref{fig:cvalpha}($a$), which presents numerically estimated $CV(x_0)$ (points) along with theoretical values (lines), the width of the domain where $CV(x_0)>1$ increases with the growing $\alpha$, i.e., for smaller $\alpha$ the domain where resetting facilitates the escape kinetics is narrower.
The set of initial conditions resulting in facilitation of the escape kinetics is further studied in Fig.~\ref{fig:cvalpha}($b$).
The bottom panel of Fig.~\ref{fig:cvalpha} shows $\xtr$ such that $CV(\xtr)=1$, i.e., $\xtr$ divides the set of initial conditions $x_0$ to such that for $|x_0| > \xtr$ the coefficient of variation is greater than 1.
From Fig.~\ref{fig:cvalpha}($b$) it clearly implies that with the increasing stability index $\alpha$ the $\xtr$ (solid line) moves towards the center of the interval and attains its asymptotic value $1/\sqrt{5}$ for $\alpha=2$, see Eq.~(\ref{eq:res-cond}), which is marked by a dashed line.
Moreover, it indicates that under action of heavy tailed noises stochastic resetting can be beneficial in narrower domain of the width $W$
\begin{equation}
    W = 2 \times (L-\xtr),
    \label{eq:width}
\end{equation}
which for $L=1$ is equal to $2(1-\xtr)$, see the dot-dashed line in Fig.~\ref{fig:cvalpha}($b$).
{The monotonous growth of $W$ with the increasing $\alpha$ should be contrasted with the (possible) non-monotonous dependence of MFPT on $\alpha$ in the absence of resetting, see Eq.~(\ref{eq:mfpt}) and \cite{szczepaniec2015escape}, or for fixed $r$, see Fig.~\ref{fig:mfptreset-lf}($a$).}
The growth of $W$ can be intuitively explained by the mechanism underlying escape dynamics.
More precisely, with decreasing $\alpha$ the dominating escape scenario is the escape via a single (discontinuous) long jump, which is less sensitive to the initial condition than escape protocol for $\alpha=2$, when the trajectories are continuous.
From Eqs.~(\ref{eq:mfpt}) -- (\ref{eq:t2}) one can also calculate the opposite, $\alpha\to 0$, limit of the MFPT: $\mathcal{T}=1$ and of the second moment: $\mathcal{T}_2=2$.
For $\alpha=0$ the Hypergeometric function in Eq.~(\ref{eq:t2}) can be replaced by unity, and the remaining integral reads $
\alpha \int_{|x_0|^2}^{L^2} \left[ t-|x_0|^2 \right]^{\frac{\alpha}{2}-1} dt
=
2\left[L^2- |x_0|^2 \right]^{\frac{\alpha}{2}}$.
Additionally, plugging $\mathcal{T}=1$ and $\mathcal{T}_2=2$ to Eq.~(\ref{eq:cv}) one gets $CV=1$, regardless of $x_0$.
Indeed, Fig.~\ref{fig:cvalpha}($a$) demonstrates that with the decreasing $\alpha$, the $CV(x_0)$ curve approaches the $CV=1$ line.
Consequently, with the decreasing $\alpha$ the $\xtr$ moves to the right, i.e., towards the absorbing boundary.
However, we are unable to reliably calculate the $\lim\limits_{\alpha\to 0} \xtr$ as numerical evaluation of analytical formulas leads to not fully controllable errors. At the same time, for $\alpha \approx 0$, stochastic simulations are unreliable. 
In overall, we are not able to provide the definitive answer whether $\xtr$ reaches edges of the interval, i.e., $\pm L$, or it stops in a finite distance to the absorbing boundary.

We finish the exploration of LF under restarting by Fig.~\ref{fig:mfptreset-lf}, which shows numerically estimated MFPTs for LF as a function of the resetting rate for various value of the stability index $\alpha$:   $\alpha\in\{0.8,1.2,1.6,1.8,2\}$.
Different panels correspond to various initial conditions: $x_0=0.5$ (top panel ($a$)) and $x_0=0.7$ (bottom panel ($b$)).
Finally, solid lines show theoretical dependence for $\alpha=2$, see Eq.~(\ref{eq:mfptreset}), while dashed lines $r=0$ asymptotics of MFPTs, i.e., $\mathcal{T}(x_0,r=0)$, see Eq.~(\ref{eq:mfpt}).
First off all, the comparison of panels ($a$) and ($b$) further corroborates that with the decreasing $\alpha$ the domain in which resetting can facilitate the escape kinetics becomes narrower.
Importantly, Fig.~\ref{fig:mfptreset-lf} clearly shows the difference between escape scenarios for L\'evy flights and Brownian motion.
For LF the dominating strategy, especially for small $\alpha$, is escape via a single long jump, while for BM the trajectories are continuous and the particle needs to approach the absorbing boundary.
This property changes the sensitivity to resetting, especially in domains where restarting hinders the escape kinetics.
For small $\alpha$ moving back to the initial condition practically does not interrupt waiting for a long jump, while for $\alpha$ close to 2 it substantially decreases the chances of escape.
Therefore, for $\alpha$ close to 2, the MFPT grows faster with increasing resetting rate.
On the other hand, the growth rate is a decaying function of the initial position, c.f., Figs.~\ref{fig:mfptreset-lf}($a$) and~\ref{fig:mfptreset-lf}($b$) for $\alpha=2$.
Moreover, from simulations we do not see facilitation of the escape kinetics due to resetting in the domain where $CV<1$.

In \cite{dybiec2017levy} similarities and differences between LF and LW have been studied.
In particular, it has been demonstrated that for LW with the power-law distribution of the jump length duration $\tau$
\begin{equation}
    f(\tau) \propto \frac{1}{\tau^{1+\alpha}}
\end{equation}
the MFPT from the $(-L,L)$ interval scales as
\begin{equation}
\label{eq:mfptLW}
    \mathcal{T}(0) \propto
    \left\{
    \begin{array}{lcc}
	 L & \mbox{for} & 0<\alpha < 1 \\
	 L^\alpha & \mbox{for} & 1<\alpha < 2 \\
	 L^2 & \mbox{for} & \alpha = 2 \\
    \end{array}
    \right.,
\end{equation}
with the half-width of the interval.
More precisely, in \cite{dybiec2017levy}, it was assumed that $v=1$ and $\tau=|\xi_\alpha|$, where $\xi_\alpha$ are independent, identically distributed random variables following a symmetric $\alpha$-stable density, see Eq.~(\ref{eq:levycf}).
The observed (asymptotic) scaling suggests that in the situation when the average jump duration/length becomes finite ($\alpha>1$) L\'evy walks display the same scaling on the interval width as L\'evy flights, see Eq.~(\ref{eq:mfpt}).
In contrast, for $\alpha<1$, the FPT for LW is bounded from below.
Namely, the first passage time $t_{\mathrm{fp}} \geqslant L/v$, which originates from the fact that the process has a finite velocity.
From this property, it implies that $\T(0) \geqslant L/v$ and thus the scaling of MFPT must differ from $ \T(0) \propto L^\alpha$ observed for LF with $\alpha < 1$.
Finally, for $\alpha = 2$, the underlying process, by means of the central limit theorem, converges to the Wiener process revealing the same scaling of the MFPT like a Brownian particle.

%
%
\section{L\'evy walks under stochastic resetting \label{sec:levywalks}}
After studying properties of LF on finite intervals under stochastic resetting, we move to examination of LW.
In the case of LW numerical simulations were conducted to investigate the regime in which stochastic resetting can be beneficial.
Similarly as in \cite{dybiec2017levy} it was assumed that $v=1$ and
$\tau=|\xi_\alpha|$, where $\xi_\alpha$ are independent, identically distributed random variables following a symmetric $\alpha$-stable density {with the scale parameter $\sigma = 1$}, see Eq.~(\ref{eq:levycf}).
{Theretofore, for LWs we eliminate two of three parameters ($v$ and $\sigma$) while we keep the $L$ parameter, see Appendix~\ref{sec:units}, as it allows for transparent examination of limiting (big number of jumps) behavior.}

For LW the first passage time density has two peaks corresponding to escape in a single long jump or a sequence of subsequent jumps towards the left or right boundary.
These peaks are located at $(L-x_0)/v$ (escape via the right boundary) or $(x_0+L)/v$ (escape via the left boundary).
Heights of peaks associated with such escapes increase with the drop in $\alpha$ and decay with the increasing interval half-width $L$.
We suspect that these peaks are one of the reasons for the emergence of differences between LF and LW, see for instance Eq.~(\ref{eq:mfptLW}) for $\alpha<1$ and discussion below Eq.~(\ref{eq:mfptLW}).

\begin{figure}
    \centering
    \includegraphics[width=0.8\textwidth]{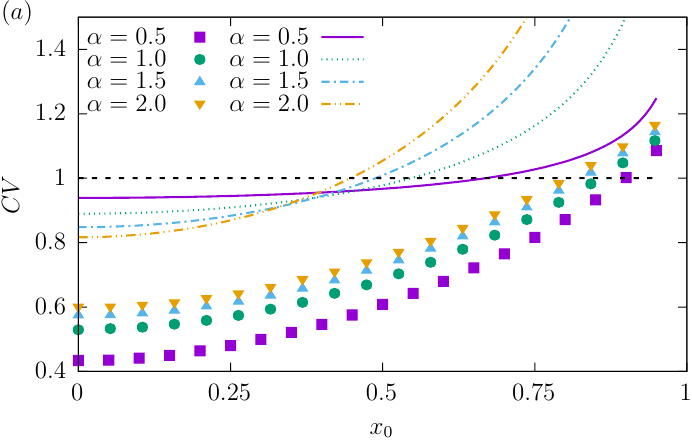}
    \includegraphics[width=0.8\textwidth]{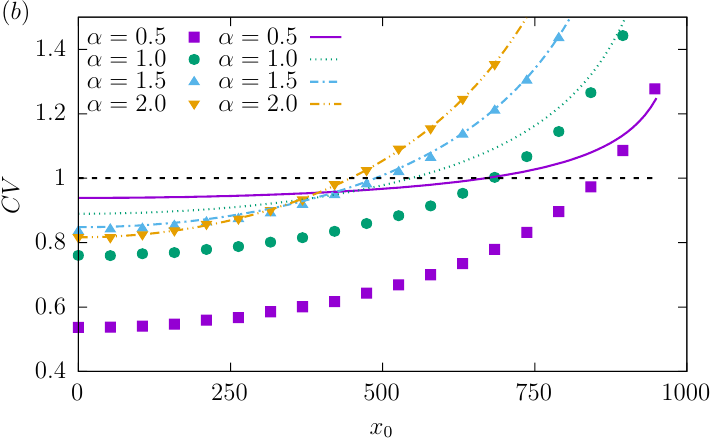}
    \caption{
	Coefficient of variation $CV$, see Eq.~(\ref{eq:cv}), as a function of the initial position $x_0$.
	Points represent $CV$ for LW (obtained through numerical simulations) and lines represent $CV$ for LF (analytical results from Eqs.~(\ref{eq:mfpt}) and (\ref{eq:t2})).
    Various panels correspond to different values of the interval half-width $L$: $L=1$ (top panel ) and $L=1000$ (bottom panel).
}
    \label{fig:CV-LW-vs-FL}
\end{figure}

The coefficient of variation $CV(x_0)$ \cite{pal2017first}, see Eq.~(\ref{eq:cv}), obtained through numerical simulations of LW onto $(-L,L)$ interval with various initial positions $x_0$ were compared to analytical results acquired for LF, see Eqs.~(\ref{eq:mfpt}) -- (\ref{eq:t2}).
We focus mainly on $1 \leqslant \alpha \leqslant 2$ case, as for $\alpha$ from that range MFPT as a function of the interval half-width $L$ for LW and LF scales in the same manner, see Eqs.~(\ref{eq:mfpt}) and~(\ref{eq:mfptLW}).
Top panel of Fig.~\ref{fig:CV-LW-vs-FL} demonstrates that for relatively small intervals half-width ($L / (v\sigma) \approx 1$), $CV$ for LF and $CV$ for LW noticeably differ, i.e., $CV$ for LW is significantly smaller than $CV$ for LF.
However, as depicted in the bottom panel of Fig.~\ref{fig:CV-LW-vs-FL}, with increasing $L$, for the same set of $\alpha$ ($\alpha \in \{ 1.5, 2\}$) the $CV$ for LW model follows the one present for LF.
The agreement originates in the fact that for large enough $L$ peaks corresponding to escape in a single jump (or sequences of consecutive jumps toward the boundary) in the first passage time distribution are small enough.

\begin{figure}
    \centering
    \includegraphics[width=0.8\textwidth]{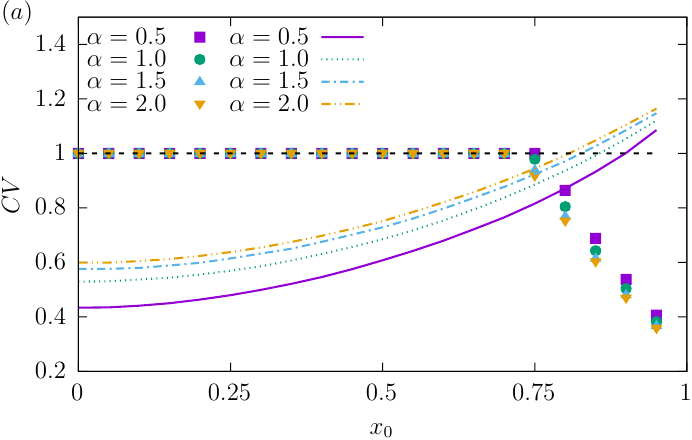}
	\includegraphics[width=0.8\textwidth]{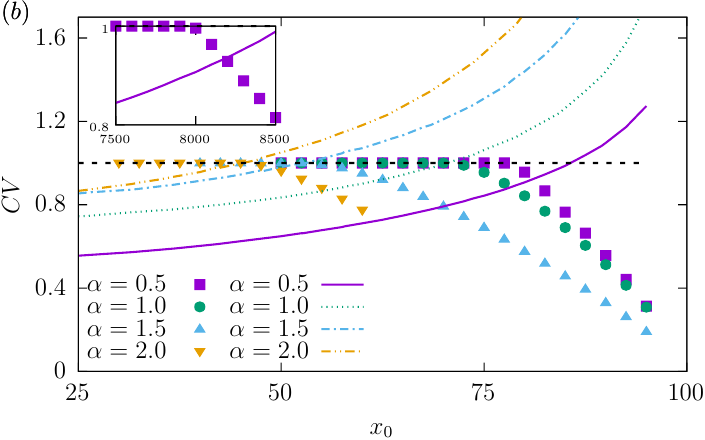}
    \caption{
    Numerically estimated (analogously like in Fig.~\ref{fig:CV-LW-vs-FL}) coefficient of variation $CV$ (lines), see Eq.~(\ref{eq:cv}), as a function of the initial position $x_0$ for LW along with the normalized ratio $\Lambda(x_0)$ (points), see Eq.~(\ref{eq:lambda}).
    Various panels correspond to different values of the interval half-width $L$: $L=1$ (top panel ), $L=100$ (bottom panel) and $L=10~000$ (inset in the bottom panel).}
    \label{fig:LW-region-helping}
\end{figure}

In the next step, the region in which stochastic resetting can be beneficial for LW was explored numerically under Poissonian (fixed rate) resetting.
The distribution of time intervals between two consecutive resets follows the exponential density $\phi(t) = r \exp(-rt)$, where $r$ ($r>0$) is the reset rate.
The efficiency of stochastic resetting can be verified by use of the normalized ratio of the minimal MFPT under stochastic resetting to its value in the absence of resetting
\begin{equation}
    \Lambda (x_0) = \frac{\min_r{\mathcal{T}(x_0, r)}}
    {\mathcal{T}(x_0, 0)},
    \label{eq:lambda}
\end{equation}
where $\mathcal{T}(x_0, r)$ stands for MFPT under resetting with reset rate $r$ and the initial position $x_0$ equivalent to the restarting point.
If stochastic resetting does not facilitate the escape kinetics $\Lambda=1$, because $\mathcal{T}(x_0, 0)$ is the minimal mean first passage time.
The decay of $\Lambda$ below one, indicates that stochastic resetting accelerates the escape kinetics.
Fig.~\ref{fig:LW-region-helping} presents the normalized ratio $\Lambda(x_0)$ (points) along with $CV(x_0)$ (lines).
For small $|x_0 / L|$ the asymmetry introduced by the initial position is not strong enough {(the restarting point is too close to the origin)} to open space for optimization of the MFPT by stochastic resetting resulting in $\Lambda(x_0)=1$.
Therefore, for small $|x_0|$, $\Lambda(x_0)=1$ not only shows the impossibility of enhancing the escape kinetics, but also introduces a visual reference level clearly demonstrating where $CV$ drops below unity.
From examination of the normalized ratio $\Lambda(x_0)$ it is possible to see easily if the drop in $\Lambda(x_0)$ coincides with the increase of the coefficient of variation $CV$ above unity.
As expected for $|x_0 / L|$ large enough stochastic resetting facilitates the escape kinetics for LW.
{For small $L$, e.g., $L=1$ (in physical units $L / (v\sigma) \approx 1$), the region where escape kinetics is accelerated by stochastic resetting differs from the one indicated by $CV>1$ criterion \cite{pal2017first}, because MFPT can be shortened even in the domain where $CV<1$, see top panel Fig.~\ref{fig:LW-region-helping}.}
This is in accordance with the fact that the condition $CV>1$ is sufficient, but not necessary, for observation of the facilitation of escape kinetics due to stochastic resetting \cite{rotbart2015michaelis,pal2019landau,pal2022inspection}.
For large enough interval half-widths $L$, the point where $\Lambda(x_0)$ drops below unity agrees with the prediction based on the $CV$ criterion for $\alpha \in \{1, 1.5, 2\}$, see bottom panel of Fig.~\ref{fig:LW-region-helping}.
However in case of $\alpha = 0.5$ the disagreement with $CV >1$ criterion persists.
{Moreover, for LWs with large $L$, the coefficient of variation agrees with the one calculated for LFs, see Fig.~\ref{fig:CV-LW-vs-FL}($b$).}

%
%
\section{Summary and conclusions\label{sec:summary}}

{
L\'evy flights and L\'evy walks constitute two paradigmatic random walks schemes generalizing the Brownian motion.
Both for LW and LF displacements are drawn from a heavy-tailed distributions with diverging variance. 
LFs are Markovian processes in which jumps occur at typical time intervals, while LWs include a
spatiotemporal coupling between jump lengths and waiting times, penalizing long jumps with long waiting times, making the process non-Markovian. 
Coupling between jump lengths and waiting times can be linear resulting in a motion with a constant velocity.  
Therefore, a random walker in LF and LW can be moving along the same turning points, however for LFs jumps are instantaneous,  while in LWs a random walker continues motion in a given direction with a constant speed for a given time after which the new direction can be generated.
As a consequence, trajectories of LWs are continuous, while for LFs they are discontinuous.
}

We have demonstrated that stochastic resetting can facilitate the escape kinetics, as measured by the mean first passage time, from a finite interval restricted by two absorbing boundaries both for LFs and LWs.
Stochastic resetting is beneficial when the initial condition, which is equivalent to the point from which the motion is restarted, sufficiently breaks the system symmetry, {i.e., $x_0$ is sufficiently far from 0 (center of the interval)}.
Under such a condition, restarting the motion anew is more efficient in bringing a particle towards the target than waiting for a particle to approach it.
Both for LFs and LWs the domain in which resetting is beneficial depends on the exponent $\alpha$ defining power-law distribution of jump lengths.
{Despite the fact that MFPT can be a non-monotonous function of stability index $\alpha$,} the width of the domain utilizing restarting is growing with the increasing $\alpha$, i.e., for lighter-tails it is wider as lighter tails change the typical escape scenario.

L\'evy flights and L\'evy walks displays the same scaling of the MFPT on the interval half-width $L$ for $1 \leqslant \alpha \leqslant 2$.
{For LF it is the exact result, while for LW it is the large $L$ asymptotics}.
Analogously, under restarting, for $1 < \alpha \leqslant 2$ with the increasing interval half-width $L$, the coefficient of variation for LWs tends {asymptotically} to the one for LFs.
Additionally, for LWs with small $L$, 
it is clearly visible that the stochastic resetting facilitates escape kinetics not only in the region where $CV>1$, but it can also accelerate the escape kinetics in situations when $CV<1$.

%
%
\section*{Acknowledgments}

We gratefully acknowledge Poland’s high-performance computing infrastructure PLGrid (HPC Centers: ACK Cyfronet AGH) for providing computer facilities and support within computational grant no.  PLG/2024/016969.
The research for this publication has been supported by a grant from the Priority Research Area DigiWorld under the Strategic Programme Excellence Initiative at Jagiellonian University.

\section*{Data availability}
The data (generated randomly using the model presented in the paper) that support the findings of this study are available from the corresponding author (B\.Z) upon reasonable request.

%
%
{

\appendix
\section{De-dimensionalization of the Langevin equation\label{sec:units}}

For LF, the $\alpha$-stable noise driven escape from $[-L,L]$ described by the Langevin equation
\begin{equation}
\frac{dx}{dt}=\xi_\alpha(t)=\sigma \xi_\alpha(\sigma=1;t)    
\label{eq:langevin-app}
\end{equation}
can be studied in dimensionless variables:
\begin{equation}
y=\frac{x}{L}\;\;\;\mbox{and}\;\;\;s=\frac{t}{T}.
\label{eq:transformation}
\end{equation}
In such variables
\begin{equation}
\frac{dx}{dt}=\frac{L}{T} \frac{dy}{ds}
\end{equation}
and
\begin{eqnarray}
\label{eq:noise}
\sigma \xi_\alpha & = & \sigma 	\frac{d}{dt} L (s T) = \sigma \frac{ds}{dt} \frac{dL(s T)}{ds}
= \frac{\sigma}{T} \frac{d}{ds }\left[ T^{\frac{1}{\alpha}} L(s) \right] \nonumber \\
& = & \sigma T^{\frac{1}{\alpha}-1} \xi_\alpha(s), 
\end{eqnarray}
In transformation of Eqs.~(\ref{eq:langevin-app}) and~(\ref{eq:noise}), we have used the following facts: (i) the scale parameter $\sigma$ can be extracted from the noise and used as the multiplicative constant \cite{janicki1994}, (ii) $\alpha$-stable noise is the formal time derivative of the $\alpha$-stable process $L(t)$, (iii) the $\alpha$-stable process is $1/\alpha$-self similar, i.e., $L(T s) = T^{\frac{1}{\alpha}} L(s)$, see \cite{samorodnitsky1994}.
In new variables, the Langevin equation takes the form
\begin{equation}
\frac{L}{T} \frac{dy}{ds} = \sigma T^{\frac{1}{\alpha}-1} \xi_\alpha(s)
\end{equation}
or
\begin{equation}
 \frac{dy}{ds} = \frac{\sigma}{L} T^{\frac{1}{\alpha}} \xi_\alpha(s).
\end{equation}
Setting $\frac{\sigma}{L} T^{\frac{1}{\alpha}}$ to unity one finds the characteristic time $T$
\begin{equation}
T=\frac{L^\alpha}{\sigma^\alpha}.
\end{equation}
In dimensionless variables the motion is described by the Langevin equation
\begin{equation}
\frac{dy}{ds} = \xi_\alpha(s)
\end{equation}
and the motion is continued as long as the particle is within  the $[-1,1]$ interval.

Incorporation of the stochastic resetting (typically) introduces another time scale associated with the interresetting intervals.
For example, for the fixed rate (Poissonian resetting) time intervals between two consecutive resets follow the exponential  density $\phi(t) = r \exp(-rt)$, where $r$ is the (fixed) reset rate.
Thus, the mean time between two consecutive restarts reads $\langle t \rangle = 1 / r$ and $r={1}/{\langle t \rangle}$.
The dimensionless reset rate $\tilde{r}$, see Eq.~(\ref{eq:transformation}), reads
$    \tilde{r}={T}/{\langle t \rangle}.$

In overall, for LFs on a bounded interval accompanied by the stochastic (Poissonian) resetting the only parameter (in addition to the stability index $\alpha$ determining tails of the jump length distribution) is the dimensionless resetting rate.
Alternatively to the above-described transformation of variables one can set $L=1$ and $\sigma=1$.
The latter approach is used in the manuscript in the part regarding L\'evy flights.

For LW the situation is not identical, but very similar to LF. LWs on a bounded interval $[-L,L]$ are characterized by three parameters: $L$ (interval half-width), $v$ (velocity of propagation) and $\sigma$ (scale of the jump time/length distribution).
Therefore, by re-scaling of the type given by Eq.~(\ref{eq:transformation}) it is not possible to remove all parameters.
After transformation of variables, a selected parameter must remain. Within current studies we have decided to keep $L$, while $v$ and $\sigma$ are set to unity.
Such a selection is motivated by the fact that it is a transparent choice allowing for study of regime in which the particle performs many jumps before leaving the domain of motion.
Subsequently, it facilitates the use of limit theorems.

}

%
%

\section*{References}

\def\url#1{}

\end{document}